\documentclass[12pt,preprint]{aastex}
\usepackage{apjfonts}
\usepackage{epsfig}
\usepackage{amsmath}
\usepackage{graphicx}
\usepackage{color}
\usepackage{float}
\usepackage{morefloats}

\def\gsim{\lower 2pt \hbox{$\, \buildrel {\scriptstyle >}\over
{\scriptstyle \sim}\,$}}
\def\lsim{\lower 2pt \hbox{$\, \buildrel {\scriptstyle <}\over
{\scriptstyle \sim}\,$}}

\begin{document}

\title{BACKGROUND MODEL FOR THE HIGH-ENERGY TELESCOPE OF \emph{INSIGHT-HXMT}}

\author{Jin-Yuan Liao\altaffilmark{1}, Shu Zhang\altaffilmark{1}, Xue-Feng Lu\altaffilmark{1}, Juan Zhang\altaffilmark{1}, Gang Li\altaffilmark{1},
Zhi Chang\altaffilmark{1}, Yu-Peng Chen\altaffilmark{1}, Ming-Yu Ge\altaffilmark{1}, Cheng-Cheng Guo\altaffilmark{1,2}, Rui Huang\altaffilmark{3}, 
Jing Jin\altaffilmark{1}, Xiao-Bo Li\altaffilmark{1}, Xu-Fang Li\altaffilmark{1}, Zheng-Wei Li\altaffilmark{1}, Cong-Zhan Liu\altaffilmark{1}, 
Fang-Jun Lu\altaffilmark{1}, Jian-Yin Nie\altaffilmark{1}, Li-Ming Song\altaffilmark{1}, Si-Fan Wang\altaffilmark{3}, Yuan You\altaffilmark{1,2}, 
Yi-Fei Zhang\altaffilmark{1}, Hai-Sheng Zhao\altaffilmark{1}, Shuang-Nan Zhang\altaffilmark{1,2,4}}
\altaffiltext{1}{Key Laboratory of Particle Astrophysics, Institute of High Energy Physics,
Chinese Academy of Sciences, Beijing 100049, China; liaojinyuan@ihep.ac.cn}
\altaffiltext{2}{University of Chinese Academy of Sciences, Chinese Academy of Sciences, Beijing 100049, China}
\altaffiltext{3}{Department of Astronomy, Tsinghua University, Beijing 100084, China}
\altaffiltext{4}{National Astronomical Observatories, Chinese Academy of Sciences, Beijing 100012, China}

\begin{abstract}
Accurate background estimation is essential for spectral and temporal analysis in astrophysics.
In this work, we construct the in-orbit background model for the High-Energy Telescope (HE) of the 
\emph{Hard X-ray Modulation Telescope} (dubbed as \emph{Insight-HXMT}). Based on the two-year blank sky
observations of \emph{Insight-HXMT}/HE, we first investigate the basic properties of the background 
and find that both the background spectral shape and intensity have long-term evolution at different 
geographical sites. The entire earth globe is then divided into small grids, each with a typical area 
of $5\times5$ square degrees in geographical coordinate system. For each grid, an empirical function 
is used to describe the long-term evolution of each channel of the background spectrum; the intensity 
of the background can be variable and a modification factor is introduced to account for this variability 
by measuring the contemporary flux of the blind detector. For a given pointing observation, the background
model is accomplished by integrating over the grids that are passed by the track of the satellite in each 
orbit. Such a background model is tested with both the blank sky observations and campaigns for observations 
of a series of celestial sources. The results show an average systematic error of $1.5\%$ for the background 
energy spectrum ($26-100$~keV) under a typical exposure of 8~ks, and $<3\%$ for background light curve 
estimation ($30-150$~keV). Therefore, the background model introduced in this paper is included in the
\emph{Insight-HXMT} software as a standard part specialized for both spectral and temporal analyses.	
\end{abstract}

\keywords{instrumentation: detectors --- space vehicles: instruments --- methods: data analysis --- X-rays: general}

\section{INTRODUCTION}
\begin{figure*}
\center{
\includegraphics[angle=0,scale=0.36]{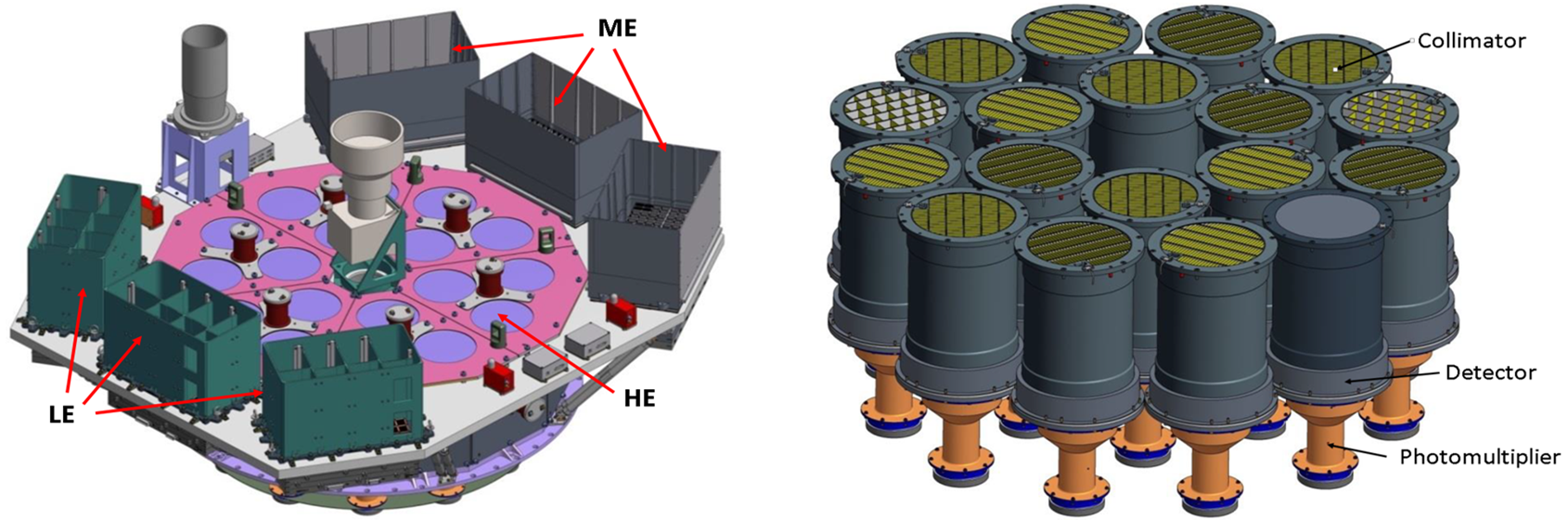}}
\caption{The main structrue of \emph{Insight-HXMT} and the HE detector modules.}
\label{Fig:HXMT_HE_structrue}
\end{figure*}

\emph{Hard X-ray Modulation Telescope} (dubbed as \emph{Insight-HXMT}) is China's first X-ray space telescope 
that was launched on June 15th, 2017. \emph{Insight-HXMT} has three telescopes with different energy bands, i.e., 
High-Energy Telescope (HE: $20-250$~keV), Medium-Energy Telescope (ME: $5-30$~keV) and Low-Energy Telescope 
(LE: $0.7-13$~keV). All the three telescopes are collimated type of telescopes with different types of detectors.
LE consists of 96 Swept Charge Devices placed in three detector boxes with a total geometrical area of 384~cm$^2$ 
(\citealt{2020_Chen_SCPMA}). ME consists of 1728 Si-PIN pixels placed in three detector boxes with a total geometrical 
area of 952~cm$^2$ (\citealt{2020_Cao_SCPMA}). HE consists of 18 NaI/CsI phoswich detectors with a total geometrical 
area of 5096~cm$^2$ (\citealt{2020_Liu_SCPMA}). The left panel of Figure~\ref{Fig:HXMT_HE_structrue} shows the main 
structrue of \emph{Insight-HXMT} and the orientations for the three telescopes. Thanks to the large detection area 
and wide detection energy range, \emph{Insight-HXMT} has carried various scientific observation tasks during the 
two years operation in orbit, e.g., the multi-band spectral and temporal analysis for pointing observation, 
wide-band and large sky scanning survey of the Galactic plane, as well as the gamma-ray observation (\citealt{2020_Zhang_SCPMA}).

The sensitivity of the telescope depends on the accuracy of the background estimation. Accurate background estimation is necessary 
for both the spectral and temporal analysis, from which the physical parameters of the observation target can be determined, such as 
the black hole spin (\citealt{1997ApJ...482L.155Z}), neutron star spin evolution (\citealt{2019RAA....19...87T}) and binary orbital 
parameters (\citealt{2012MNRAS.421.3550Z}). Up to now, dozens of X-ray space telescopes have been operated in orbit, and the background 
estimation methods are different from each other. For a focusing telescope, the X-ray sources can be directly imaged, so the 
background can be obtained from the surrounding area of the sources in the image, e.g., {\sl XMM-Newton} and {\sl Chandra} 
(\citealt{2001A&A...365L..18S}; \citealt{2001A&A...365L..27T}; \citealt{2003SPIE.4851...28G}). For a collimating telescope, 
the background can be obtained from the on-off observation as {\sl RXTE}/HEXTE (\citealt{1998ApJ...496..538R};
\citealt{2014ApJ...794...73G}, 2016) and {\sl BeppoSAX}/PDS (Frontera et al. 1997a, 1997b).
In addition, the background can also be calculated from the background model with a series of parameters of the instrument and 
the space environment as {\sl Suzaku}/HXD (\citealt{2009PASJ...61S..17F}).

As a collimated type of telescope, \emph{Insight-HXMT} must develop its own background estimation method due to its unique design.
Some of the background of all the three telescopes of \emph{Insight-HXMT} are generated by the interaction between the satellite 
and space particles. For the three telescopes of \emph{Insight-HXMT}, we construct the background models based on the 
blank sky observations in two years and make detailed systematic error analyses to their background models, respectively. 
Each telescope has different background characteristics and different background estimation method. In this paper, 
we focus on the background of the high-energy telescope of \emph{Insight-HXMT}. The background of the medium-energy and low-energy 
telescopes will be described in another paper.

This paper is organized as follows. We give an overview of the instrument and background of \emph{Insight-HXMT}/HE in Section 2. 
In Section 3, we describe the method to estimate the background. In Section 4, we show the model check and analysis of systematic 
error of the background estimation. The discussion and summary appear in Sections 5 and 6, respectively.

\section{INSTRUMENT AND BACKGROUND}
\begin{figure*}
\center{
\includegraphics[angle=0,scale=1.00]{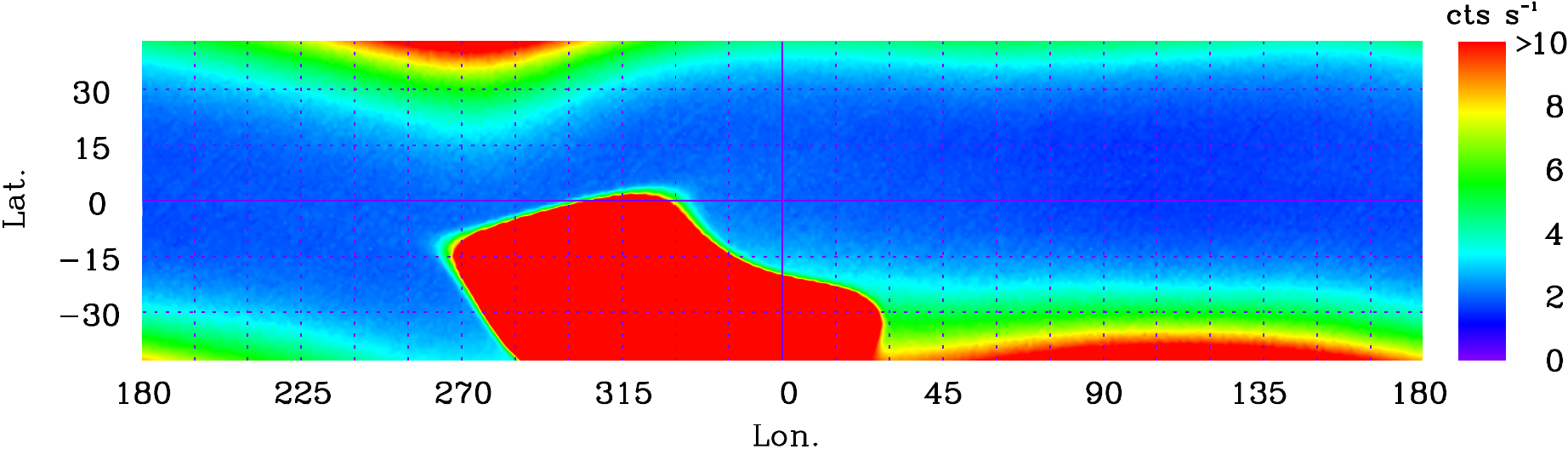}}
\caption{Geographical distribution of PM count rate.}
\label{Fig:PM_map}
\end{figure*}

\emph{Insight-HXMT} operates in a quasi-circular low-earth orbit with an attitude of $\sim$550~km and an inclination of 
$\sim$43$^\circ$, where the space environment is complex and exists various particles (e.g., Alcaraz et al. 2000a, 2000b). 
Cosmic-ray protons (CRP) are dominant, but electrons, neutrons, cosmic X-ray background (CXB), and the gamma-rays reflected 
from the earth (Albedo) also can not be neglected. The satellite platform and the payload interact with the particles, 
contributing to the background of \emph{Insight-HXMT} (\citealt{2009ChA&A..33..333L}; \citealt{2015Ap&SS.360...47X}). 
The Particle Monitor (PM) onboard \emph{Insight-HXMT} can give the geographical distribution of protons ($>20$~MeV) and 
electrons ($>1.5$~MeV) in the orbit of \emph{Insight-HXMT} (\citealt{2020_Lu_JHEAp}). As shown in Figure~\ref{Fig:PM_map}, 
PM count rate is low ($<3$~cts~s$^{-1}$) in the vicinity of the equator and is high ($>10$~cts~s$^{-1}$) in high-latitude region; 
in the South Atlantic Anomaly (SAA), PM count rate is much higher than that outside SAA. All the PM results are similar with the 
result of Radiation-Belt Monitor (\citealt{2011PASJ...63S.635S}), despite the altitude and the orbital inclination are moderately 
different with each other.
\subsection{Instrument}
\begin{figure*}
\center{
\includegraphics[angle=0,scale=1.00]{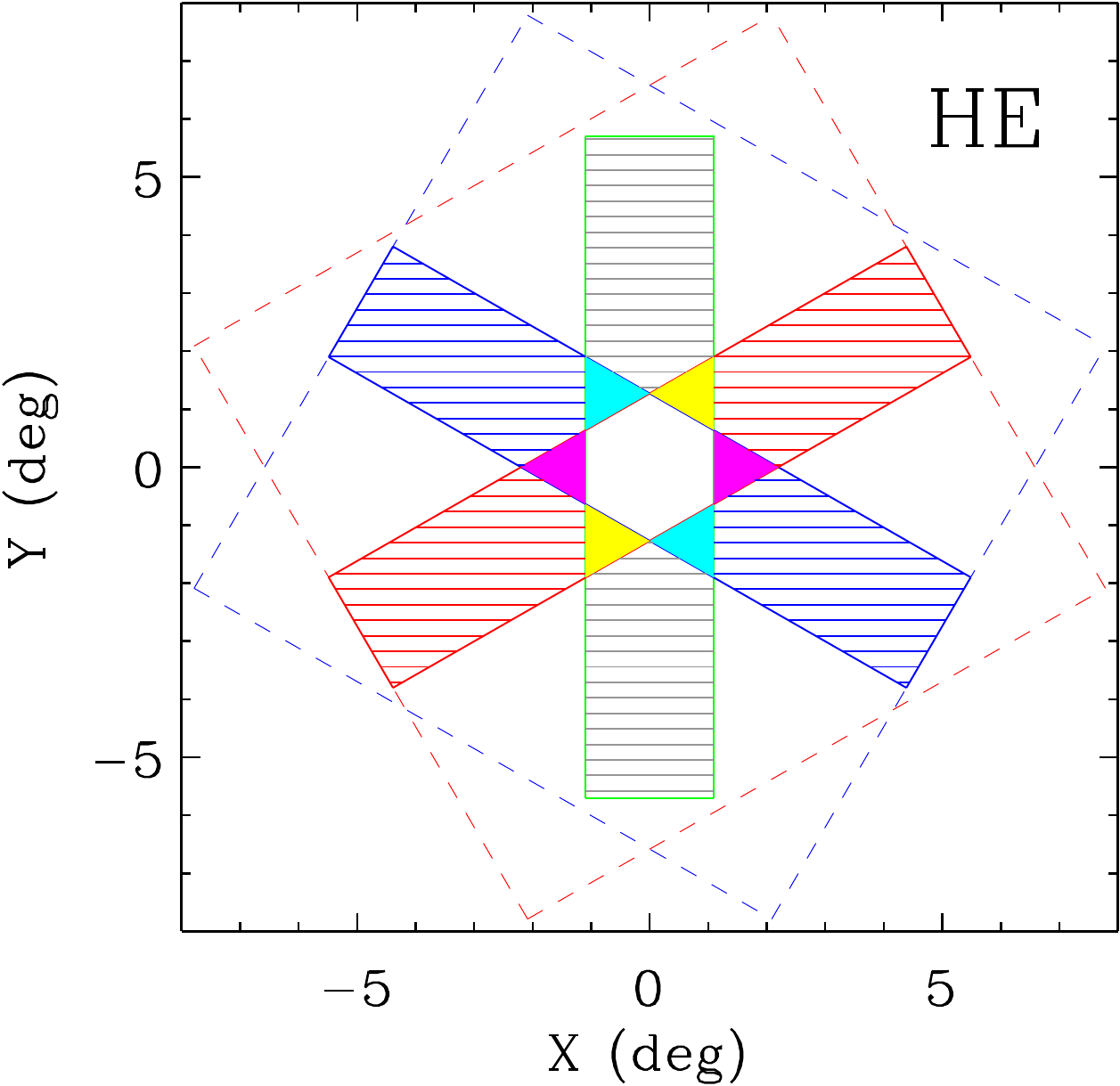}}
\caption{FOVs of the \emph{Insight-HXMT}/HE detectors. Three rectangular shadow shown in red, blue and green are three small FOVs 
(FWHM: 1$^\circ.1\times5^\circ$.7); and the square shown in red and blue are two large FOVs (FWHM: 5$^\circ.7\times5^\circ$.7)}
\label{Fig:HE_FOV}
\end{figure*}

As described in \citet{2020_Liu_SCPMA}, \emph{Insight-HXMT}/HE has 18 NaI/CsI phoswich detectors (Figure~\ref{Fig:HXMT_HE_structrue}), 
15 of which belong to three small FOVs (FWHM: 1$^\circ.1\times5^\circ$.7) with the orientation offset by 60$^\circ$ (each direction 
contain 5 detectors) and two detectors belong to two large FOVs (FWHM: 5$^\circ.7\times5^\circ$.7) with the orientation offset by 
60$^\circ$ (Figure~\ref{Fig:HE_FOV}). In addition, there is also a detector with a blind FOV (collimator covered by a tantalum sheet). 
In orbit, the interactions between the particles and the satellite generate background with different time scales. From the simulation 
on ground, the HE background is mainly dominated by the particle background with short time scale and the activation background with 
long time scale (\citealt{2009ChA&A..33..333L}; \citealt{2015Ap&SS.360...47X}). \emph{Insight-HXMT}/HE has the design with active background 
shielding, i.e., the instrument is surround by 18 plastic scintillator detectors that are also called anti-coincidence detectors (ACD). 
When a high-energy proton passing through, it will not only deposit energy in NaI crystal, but also makes a signal in ACD at the same time 
(within 1~$\mu$s). These events will be marked in the Event File (the proportion $\sim30\%$ for blank sky observation) and will be excluded 
when we do the data analysis. For the data analysis of a pointing observation, a part of the background is removed by the active shielding 
technique. In this paper, we only focus on the remaining background after the anti-coincidence.

\subsection{Light Curve, Spectra, and Background Map}
\begin{figure*}
\center{
\includegraphics[angle=0,scale=1.00]{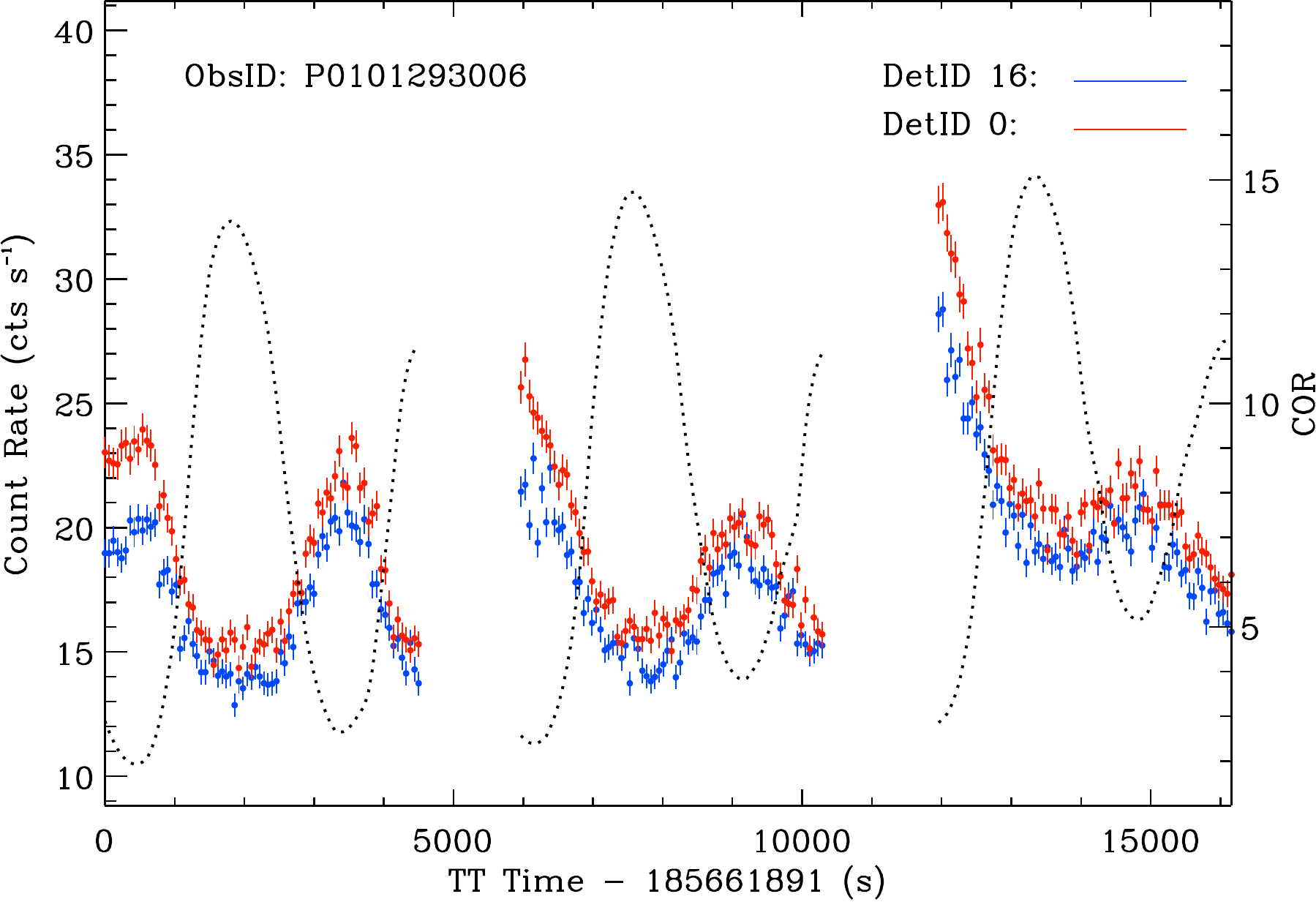}}
\caption{Observational background light curves ($30-150$~keV and ${\rm ObsID}={\rm P}0101293006$). 
The observational light curve of $DetID=16~\&~0$ are shown in blue and red, and the COR curve are shown with the dotted line.}
\label{Fig:bkg_light_curve}
\end{figure*}

\begin{figure*}
\center{
\includegraphics[angle=0,scale=1.00]{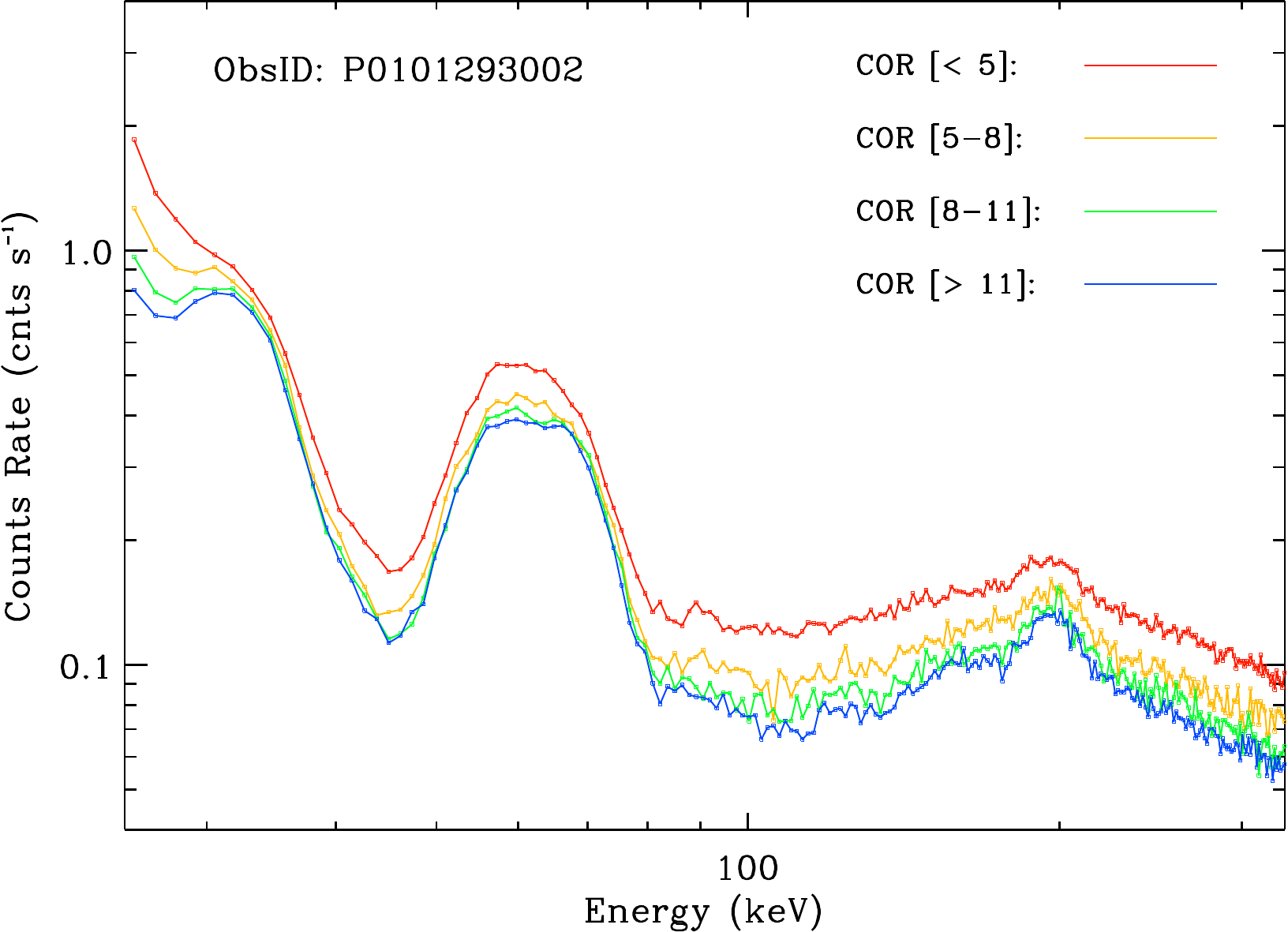}}
\caption{Observational background spectra with different COR ranges (${\rm ObsID}={\rm P}0101293002$).}
\label{Fig:bkg_spectra}
\end{figure*}

\begin{figure*}
\center{
\includegraphics[angle=0,scale=1.00]{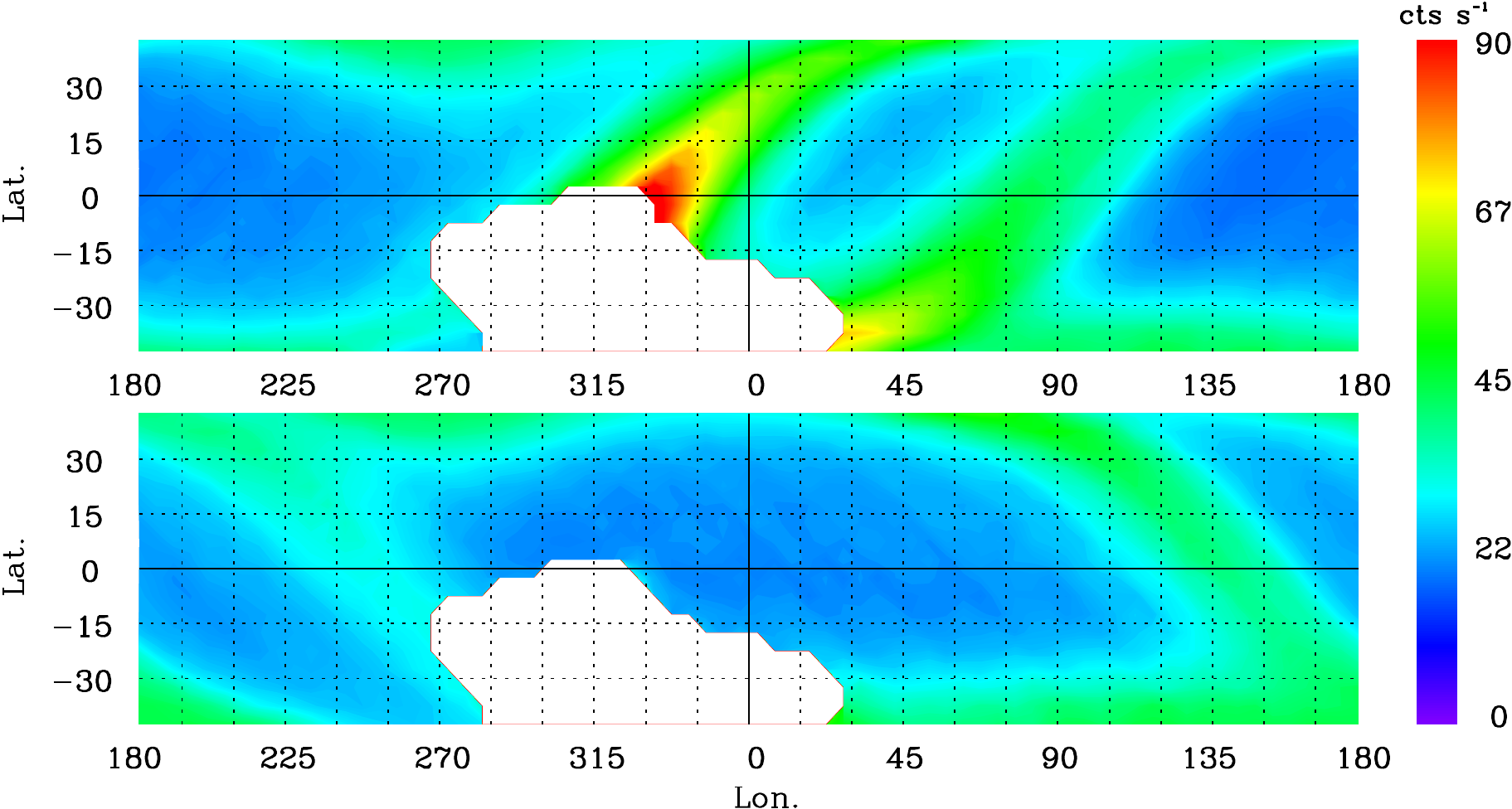}}
\caption{Geographical distribution of the background intensity (${\rm DetID}=0$ \& $20-250$~keV).
The top and bottom panels show the results as \emph{Insight-HXMT} in ascend and descend orbital phase, respectively.}
\label{Fig:bkg_map}
\end{figure*}

  The variety of the light-curves of HE background is remarkable, which has two significant features. The first is the modulation of the 
geographical location. The orbital inclination of \emph{Insight-HXMT} is 43$^\circ$, hence the satellite passes through the high latitude 
region and the equatorial region twice in one orbital period. As shown in \citet{2009PASJ...61S..17F}, the geomagnetic cut-off rigidity (COR) 
is usually low relatively in the high latitude region and is high in the equatorial region. As the charged particle flux is inversely 
proportional to COR, thus the background intensity also has a significant anti-correlation with COR (Figure~\ref{Fig:bkg_light_curve}). 
Second, the background rises rapidly when the satellite passes through SAA (No data when \emph{Insight-HXMT} is shutdown for protection) 
and decreases gradually after the satellite leaves SAA, indicating the decay rate exceeds the activation rate. The most obvious feature of 
the HE background spectrum is the activation emission lines (e.g., 33~keV, 65~keV, 191~keV). Although the blind detectors cannot accept X-ray 
photons less than 200~keV, however, as the proportion of CXB is very low in HE background, the spectrum of the blind detector is consistent 
with that of the non-blind detector. The background spectrum consists of multiple components, and each component has a different spectral 
shape (\citealt{2009ChA&A..33..333L}; \citealt{2015Ap&SS.360...47X}). As the proportion of each component changes with COR, the background 
spectrum also varies with COR (Figure~\ref{Fig:bkg_spectra}). The HE telescope is severely activated every time it passes through SAA. In 
Figure~\ref{Fig:bkg_map}, the upper and lower panels are the background maps for the satellite at the ascend and descend phases of a small 
FOV detector. For \emph{Insight-HXMT}, the flight direction at the ascend phase is from south to north, and is from north to south at the 
descend phase. It can be seen that the intensity of the background is very different between the two panels in the same geographical position. 
This is mainly due to the difference of the degree of activation decay. In addition, because the proportions of the activating components are 
different, the spectra (both the shape and intensity) in the same geographical location are also different.

\subsection{Data Reduction}
\begin{figure*}
\center{
\includegraphics[angle=0,scale=1.00]{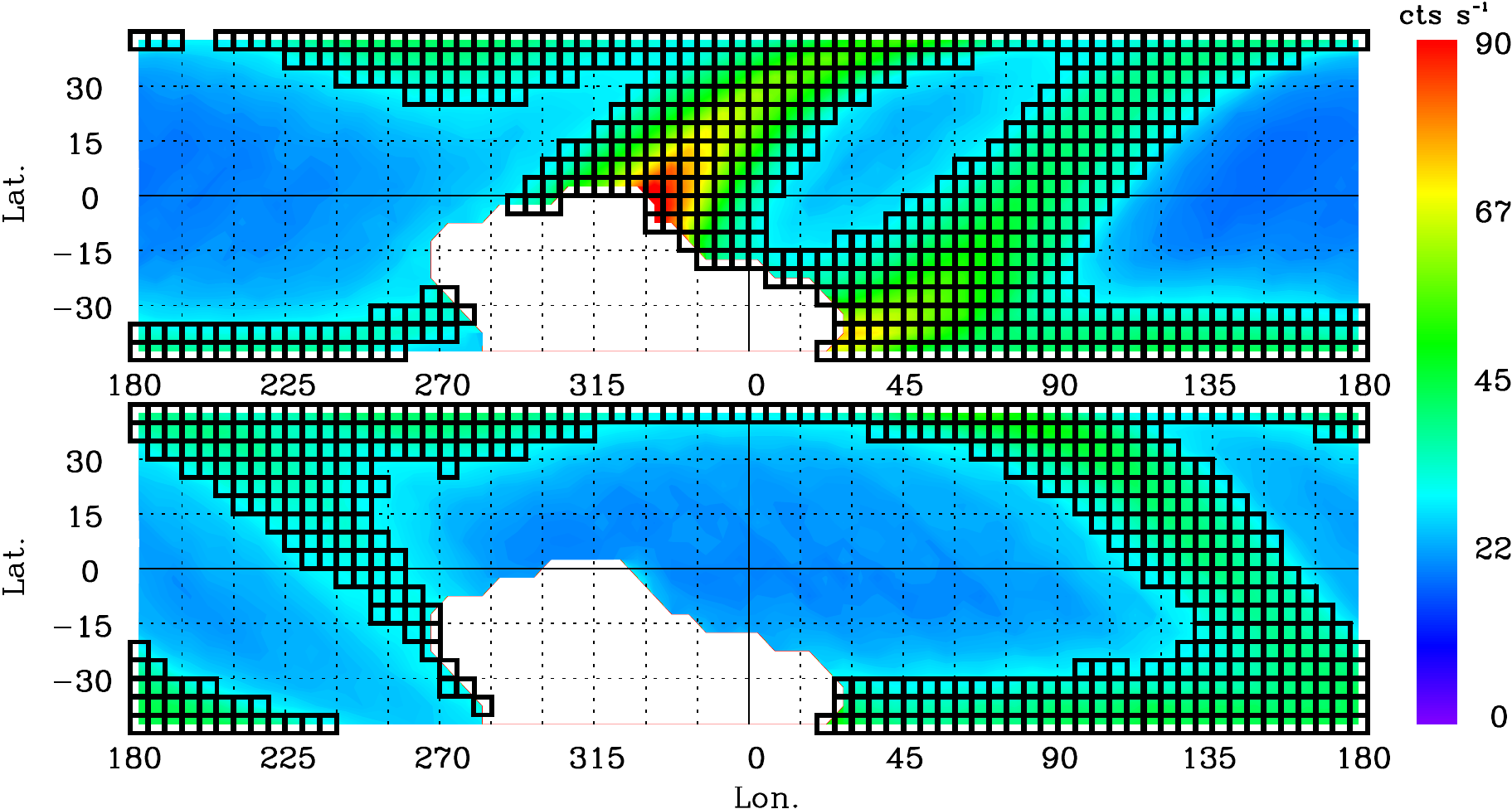}}
\caption{Geographical region excluded in GTI selection. The region marked by black squares is excluded from the GTI (top: ascend phase; bottom: descend phase)}
\label{Fig:GTI_region}
\end{figure*}

  Since the earth's atmosphere can interact with the X-ray photons and distort the intrinsic spectrum of the target, we choose ELV~$>10$ for 
pointing observations. When the satellite is at high latitude region, COR is small and the background intensity will be very high, so we 
choose COR~$>8$. Since the HE background is dominated by the activation component, the background of HE will rise rapidly every time as the 
satellite passes through SAA and then gradually decreases. Combined with the characteristics of \emph{Insight-HXMT}, the data with the time 
300~s before and after SAA is also excluded. SAA is very large and sometimes the satellite will be severely activated when the satellite 
passes through SAA. Even a thousand seconds after the satellite passes through SAA, the background is still very high. Due to the relatively 
stable space environment, the relationship between HE background and geographical location is also stable, thus we can obtain where HE is 
affected mostly. As shown in Figure~\ref{Fig:GTI_region}, the region with the strongest activation background is marked with the purple 
squares. In order to make the background estimation more accurate, we currently exclude these areas from Good Time Interval (GTI). 
In the future, after the further study of the background, we will gradually reduce the region until it is completely included in the GTI.

\section{BACKGROUND MODELING}
\subsection{Modeling Background Spectra}
\begin{figure*}
\center{
\includegraphics[angle=0,scale=1.00]{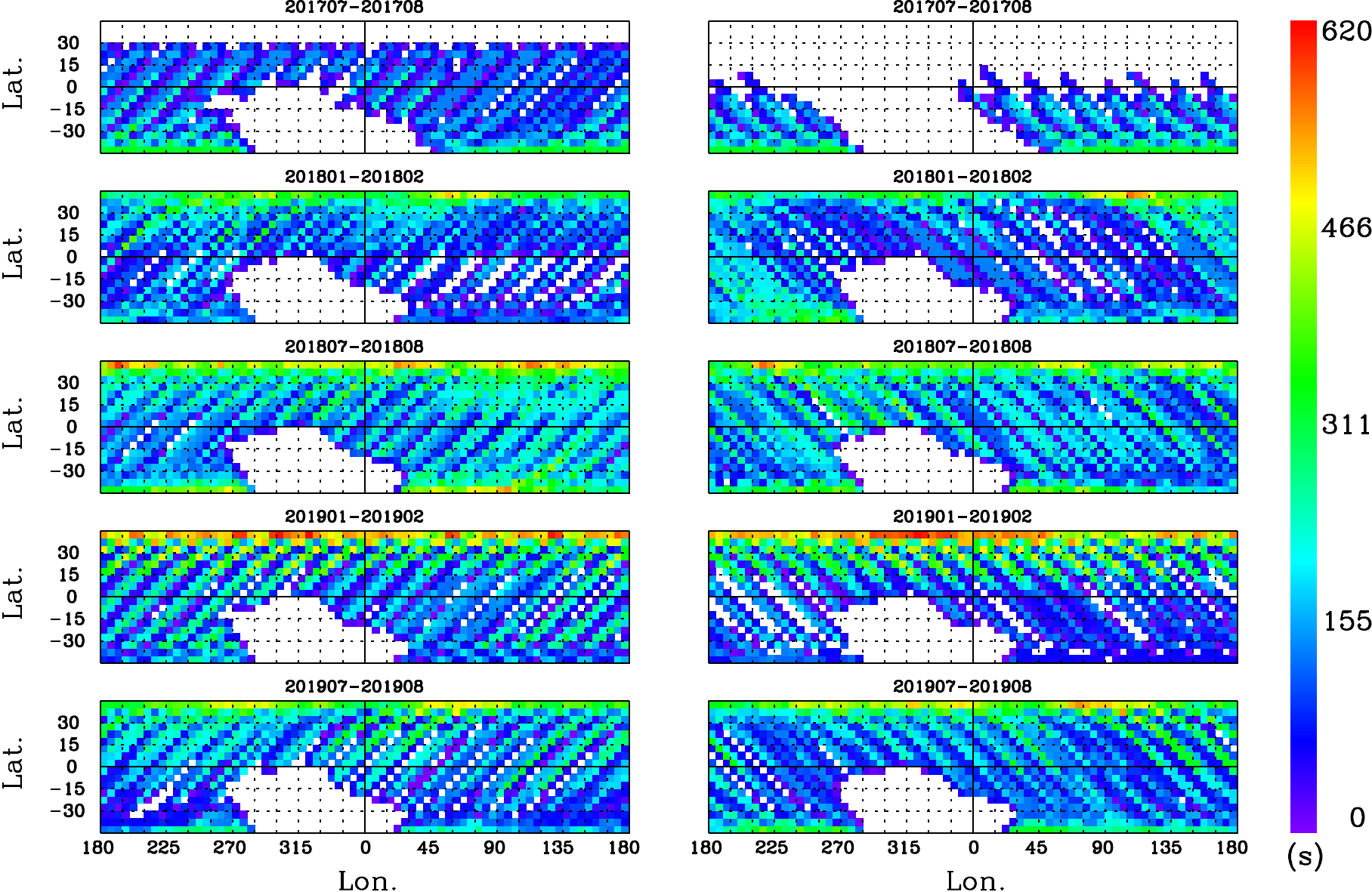}}
\caption{Coverage of geographical location by the blank sky observations in five periods that contain two months.
The left five panels are for the flight direction at the ascend phase, 
those for the flight direction at the descend phase are shown in the right five panels.}
\label{Fig:Coverage_blank_sky}
\end{figure*}

  As described in Section 2, the space environment in the orbit of \emph{Insight-HXMT} is very complicated. The satellite interacts with 
various particles (cosmic protons, albedo Gamma-rays, and cosmic X-ray backgrounds), resulting in various background components with different 
time-scales. Since the space environment is basically stable, each background component is related to the geographical location closely and 
the proportion of each background component varies with the geographic location. Therefore, we must study the characteristics of the 
background in detail and then build the database. Up to December 12th, 2019, \emph{Insight-HXMT} has made more than 200 blank sky observations 
(effective exposure $\sim3.5$~Ms), which are used to build the background database. The coverage of geographical location by the blank sky 
observations in every two months are shown in Figure~\ref{Fig:Coverage_blank_sky}. The reflected X-ray from the earth can enter the aperture 
and be detected by HE, however, the earth will never appear in the field of view during the effective observational time for the pointing 
observation. Therefore, we do not use the data with earth occultation to build the background database in order to improve the accuracy of the 
background model.
  
\begin{figure*}
\center{
\includegraphics[angle=0,scale=1.00]{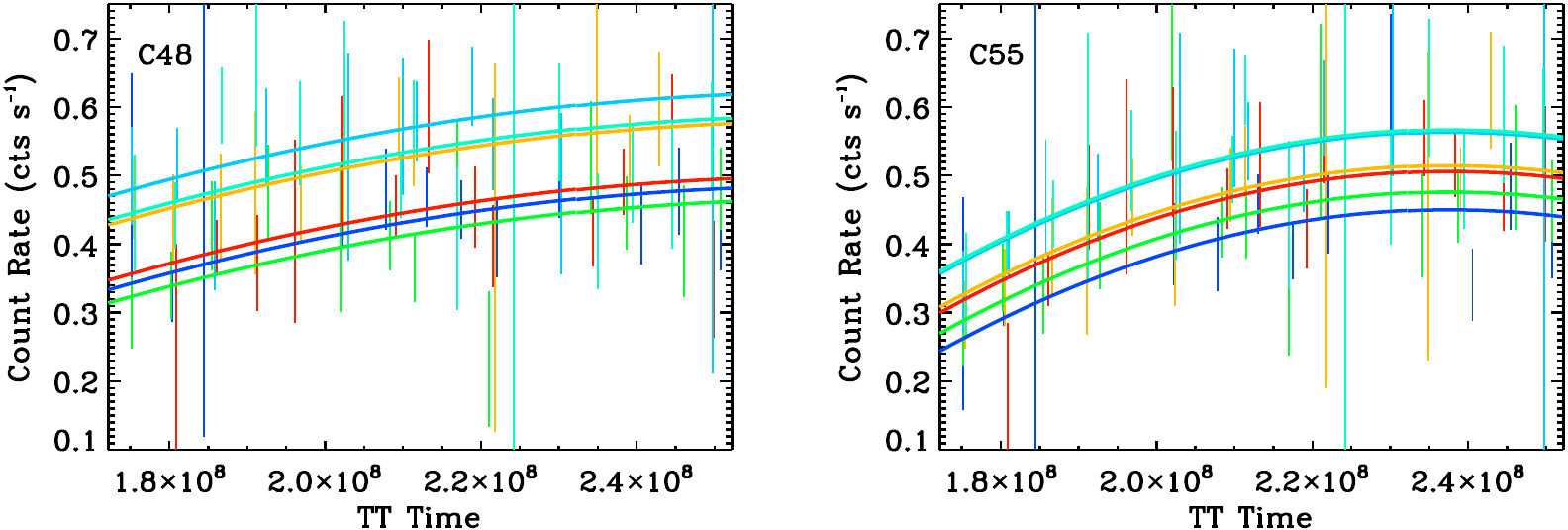}}
\caption{Long-term light curves at six geographical locations (shown with different colors) of the 48th (left) and 55th (right) channel.}
\label{Fig:long_term_fit}
\end{figure*}

  Despite the background has a variety of components, the relative proportions of the background components are stable during the same period 
and in the same geographical location. On the long time scale, the background at the same geographical location increases with time and the 
increase rate of different channel is different (Figure~\ref{Fig:long_term_fit}), thus the shape of the background spectrum also changes with 
time. The long-term evolution of the background spectra at different geographic locations is consistent with each other, because the long-term 
decay of the activation background is independent of geographic location.
  
  We build a database of all the 18 HE detectors, each of which contains the parameters: geographic location ($lon$: longitude \& $lat$: 
latitude), flight direction ($fd$: ascend \& descend), channel ($c$), observation time ($t$). The orbital inclination of \emph{Insight-HXMT} 
is 43$^\circ$, covering a geographic location of $\sim360\times43\times2$ square degrees. We divide the geographic location into 72$\times$18 
grids (5$\times$5 square degrees per grid). For every flight direction, the background spectra in each grid ($f_{\rm g}$) can be described by 
an empirical function. Here, a quadratic polynomial is used as:
\begin{equation}
f_{\rm g}(c;t,lon,lat,fd) = A(c;fd)t^{2} + B(c;fd)t + C(c;lon,lat,fd).
\end{equation} 
  In a flight direction, the parameters $A$ and $B$ only depend on the channel. The parameter $C$ depends on not only the channel, but also 
the geographical location. That is, in the same geographical location and flight direction, the intensity of the background spectrum in each 
channel can be described by a quadratic function, respectively. The parameters of each channel are different, thus the shape of the background 
spectra also evolve with time. For each channel of a detector in the ascend phase, we fit the long-term evolution with quadratic function at 
$72\times18=1296$ geographical locations, i.e., 1296 geographical locations have the same $A$ and $B$, and different $C$ (a total of 1298 
fitting parameters). After the empirical function fitting, the background database has been parameterized, and then the expected value of the 
background spectrum in any geographical location $f_{\rm g}(c;t,lon,lat,fd)$ can be obtained by specifying only $t$, $lon$, $lat$, and $fd$. 
For a pointing observation, we can obtain the orbital parameters (geographical locations and flight directions), and then the orbit can be 
divided into several grids ($lon, lat, fd$). The exposure of every grid ($T_{\rm g}$) is also obtained, thus the expected background spectrum of every grid can be obtained by equation 1. The expected background spectrum of a pointing observation can be calculated by adding all the expected background spectrum of all the grids,
\begin{equation}
S_{\rm exp}(c;t) = \frac{\sum_{lon,lat,fd}f_{\rm g}(c;t,lon,lat,fd)T_{\rm g}(lon,lat,fd)}{\sum_{lon,lat,fd}T_{\rm g}(lon,lat,fd)}.
\end{equation}
  The space environment is generally stable, however, it still has small fluctuations (space particle proportion, intensity, spectral shape). 
The expected value cannot describe the background of a pointing observation accurately. Therefore, the expected value in each parameter grid 
must be corrected, where the blind detector will be used to do the correction. The blind detector is the unique design of \emph{Insight-HXMT}. 
As described in Section 2, the blind detector cannot receive X-ray photons in the direction of the collimator aperture as the collimator is 
covered by a tantalum sheet. For one grids of a pointing observation, the expected background spectrum of the blind detector 
$f_{\rm exp}(c;t,lon,lat,fd|{\rm BD})$ can be obtained by equation 1, as well as the observed spectrum $f_{\rm obs}(c;t,lon,lat,fd| {\rm BD})$ 
of this observation, thus the correction factor can be expressed as:
\begin{equation}
\mathcal{F}_{\rm g}(c;t,lon,lat,fd) = \frac{f_{\rm obs}(c;t,lon,lat,fd|{\rm BD})}{f_{\rm exp}(c;t,lon,lat,fd|{\rm BD})},
\end{equation}
and then, we obtain
\begin{equation}
S_{\rm bkg}(c;t) = \frac{\sum_{lon,lat,fd}f_{\rm g}(c;t,lon,lat,fd)T_{\rm g}(lon,lat,fd)\mathcal{F}_{\rm g}(c;t,lon,lat,fd)}{\sum_{lon,lat,fd}T_{\rm g}(lon,lat,fd)},
\end{equation}
where $S_{\rm bkg}(c;t)$ is the at the estimated background spectra $S_{\rm bkg}(c)$ at observation time $t$.

  The calculation process can be summarized as follows:
  
  (1) Build the background observation database, including geographic location, flight direction, channel, and the observation time;
  
  (2) Fit the long-term evolution of the background in database with the empirical function (quadratic polynomial), 
  in order to obtain the background of each detector for any observation time;
  
  (3) For a pointing observation, obtain the exposure of each detector in every grids of the background database during the observation;
  
  (4) Obtain the expected background spectrum in each grid of the blind detector as well as the observation spectrum, and then obtain the 
  correction factor;
  
  (5) Correct the expected spectrum in each grid of other non-blind detectors;
  
  (6) Integrate the corrected background spectrum of the non-blind detectors in all the grids to obtain the final background spectrum of the 
  pointing observation.

\subsection{Modeling Background Light Curve}
The estimation method of the background light curve is consistent with that of the background spectra in principle, 
where the background database and blind detector are also used. The specific steps are as follows:

(1) We obtain the observational light curve of the blind detector (BD) within a specified energy band $\Delta E$: $L_{\rm obs}(t;\Delta E|{\rm BD})$;

(2) The ratio of source detector (SD) to blind detector at every time bin of the light curve are obtained as
\begin{equation}
\mathcal{R}(t;\Delta E) = \frac{L_{\rm exp}(t;\Delta E|{\rm SD})}{L_{\rm exp}(t;\Delta E|{\rm BD})},
\end{equation}
where $L_{\rm exp}(t;\Delta E|{\rm SD})$ and $L_{\rm exp}(t;\Delta E|{\rm BD})$ are the expected background light curves 
of the source detector (SD, small and large FOVs) and blind detector that can be calculated from the background database;

(3) The background light curve of the source detector at specified energy band $\Delta E$ can be obtained by
\begin{equation}
L(t|{\rm SD}) = L_{\rm obs}(t;\Delta E|{\rm BD})\mathcal{R}(t;\Delta E).
\end{equation}

There is a technical detail that worth pointing out.
For a light curve with small time bin (e.g., $\rm T_{bin} = 1~{\rm s}$), the counts at each time bin are too low to do the statistical analysis.
As shown by the on-ground simulation, the time scale of the variability of the HE background is always very long ($>100$~s) that is also confirmed by observation,
we first generate a background light curve with $\rm T_{bin} = 16~{\rm s}$, 
and then the background light curve with shorter time bin can be obtained by the interpolation.

\section{MODEL TEST AND ANALYSIS OF SYSTEMATIC ERROR}
\subsection{Test for Background Spectra}
\begin{figure*}
\center{
\includegraphics[angle=0,scale=1.00]{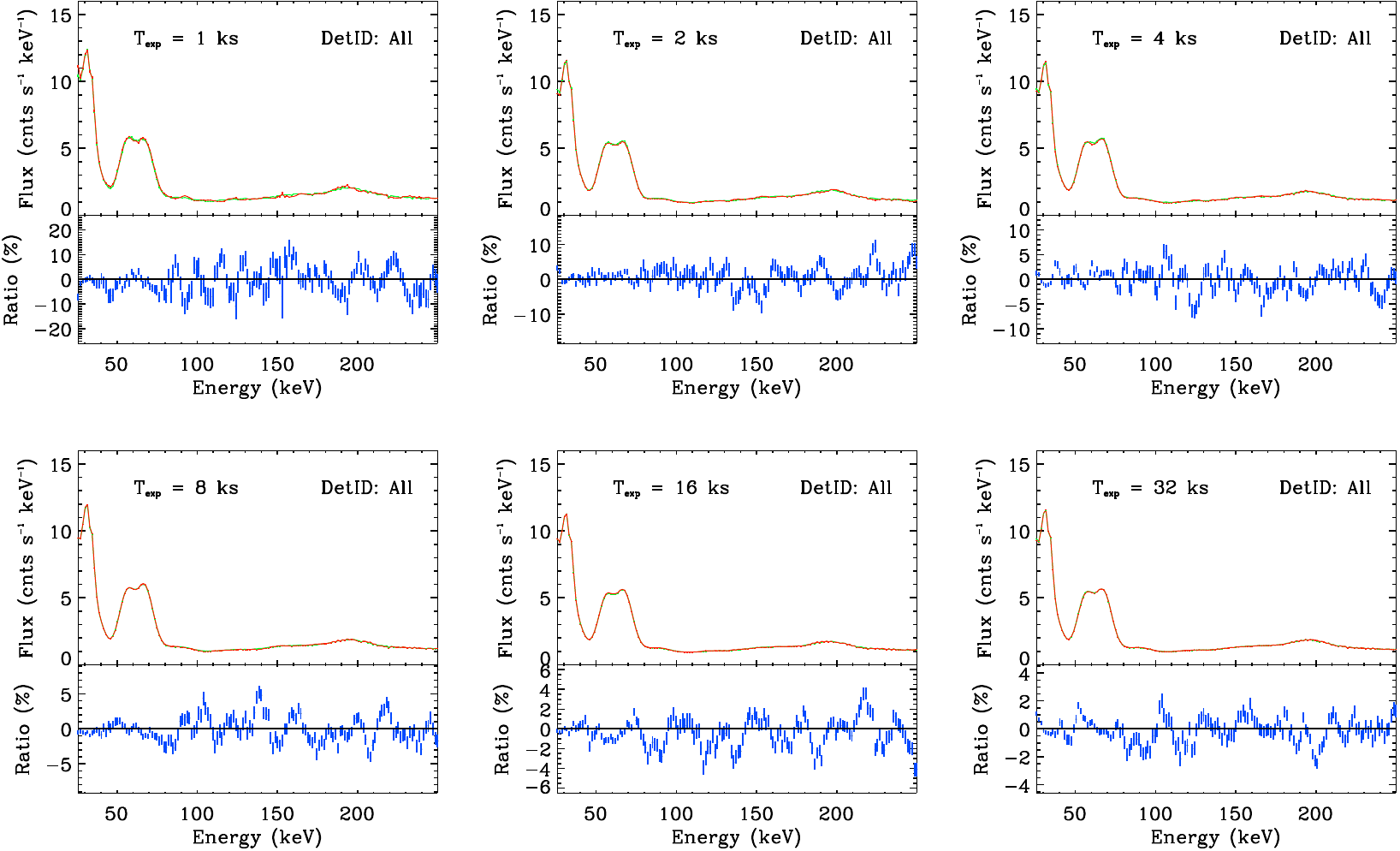}}
\caption{Background spectra test with different exposures. Top: green is observational spectra and blue is estimation; bottom: Ratio of the residual and the data.}
\label{Fig:bkg_spec_test}
\end{figure*}

With the background model described in Section 3, the background spectra of any pointing observation can be obtained.
We use the observation of the blank sky to test the background model.
In order to ensure the independence of the model and test data, all the blank sky observations with effective exposure $\sim3.5$~Ms
are divided into two parts, i.e., the first part ($\sim3$~Ms) is used to construct the background model, and the second part ($\sim0.5$~Ms)
is used as the test data. 
Figure~\ref{Fig:bkg_spec_test} is an example showing the background spectra estimation of a blank sky observation with 
exposures $T_{\rm exp}=1~{\rm ks}$, 2~ks, 4~ks, 8~ks, 16~ks, and 32~ks.
In order to test the estimated background spectra with a specific exposure (e.g., $T_{\rm exp}=1~{\rm ks}$), 
all the data will be combined first, and then regrouped into the sub-observations with the specific exposure. We take $T_{\rm exp}=1~{\rm ks}$ as an example 
to show the calculation process of the systematic uncertainty of the estimated background spectra, the detailed steps are shown as follows:

(1) Re-group all the blank sky observations for model test into $N$ sub-observations with $T_{\rm exp}=1~{\rm ks}$.

(2) Obtain the observational spectra of these sub-observations $S_{i, \rm obs}(c)$ and the background spectrum $S_{i, \rm bkg}(c)$ calculated 
by the background model, where $i=1,2,...,N$.

(3) Make a comparison between $S_{\rm obs}(c)$ and $S_{\rm bkg}(c)$ to obtain the residuals $R(c)=S_{\rm obs}(c)-S_{\rm bkg}(c)$.

(4) Calculate the mean value and intrinsic dispersion of $R(c)$, i.e., $R_{\rm m}$ and $D_{\rm in}$, by solving the following equation
\begin{equation}
\sum_{i=1}^N \frac{(R_{i}-R_{\rm m})^2}{D_{i}^2} = N - 1,
\end{equation}
where
\begin{equation}
    D_{i}^{2} = D_{\rm in}^{2} + \sigma_{i}^{2},
\end{equation}
\begin{equation}
    R_{\rm m} = \sum_{i=1}^N R_{i}\times w_{i}, \quad w_{i} = \frac{\frac{1}{D_{i}^{2}}}{\sum_{i=1}^N \frac{1}{D_{i}^{2}}},
\end{equation}
where $R_i$ refers to the $R(c)$ of every observation, $\sigma_{i}$ the statistic errors of $R_i$. 
$R_{\rm m}$ and $D_{\rm in}$ can be considered as the bias ($B$) and the systematic error ($\sigma_{\rm sys}$) of the background estimation.

\begin{figure*}
\center{
\includegraphics[angle=0,scale=1.00]{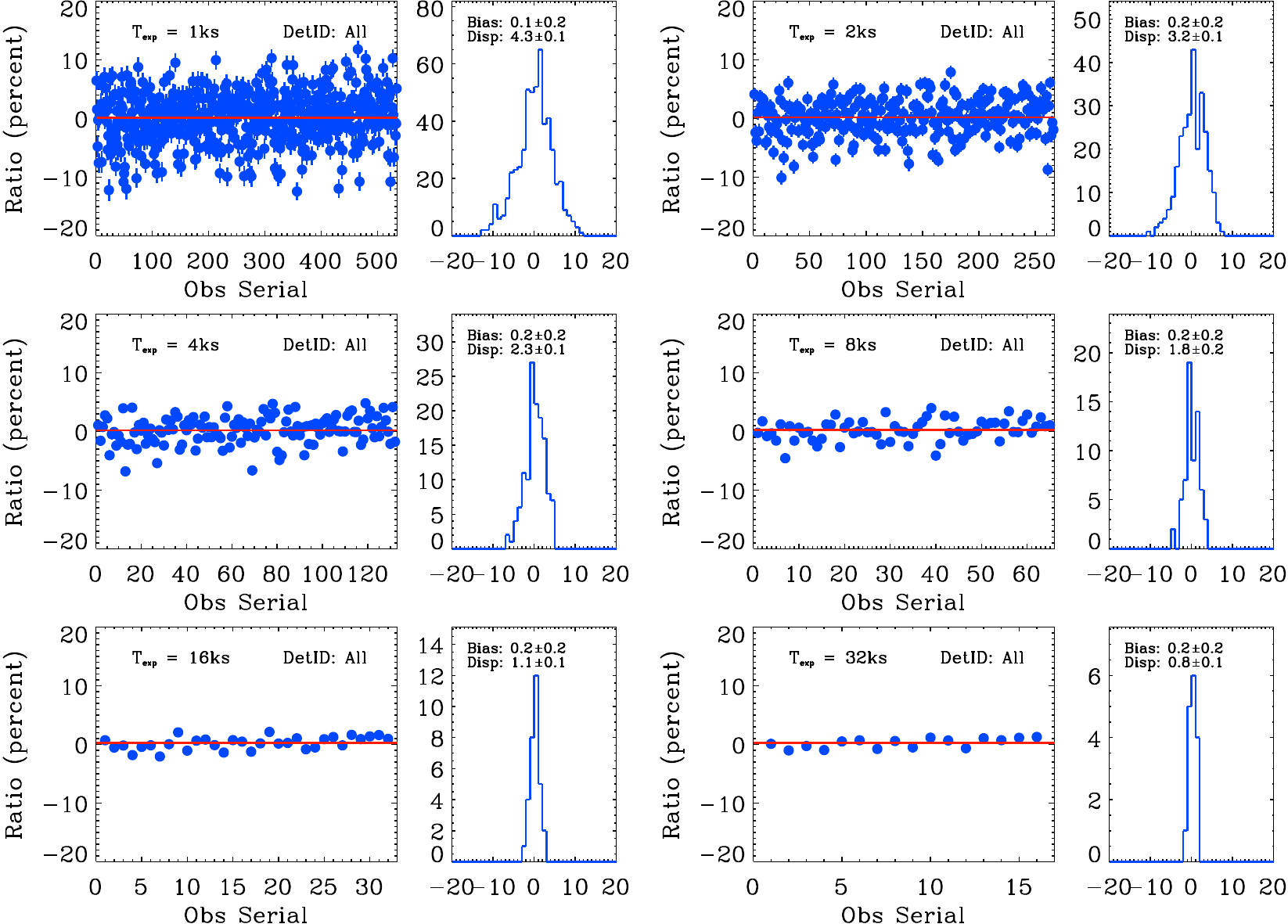}}
\caption{Distributions of the residuals in the background spectra estimation in the 40th channel with exposures 1~ks, 2~ks, 4~ks, 8~ks, 16~ks and 32~ks.
In each panel, the broadening of the histogram is the dispersion of the residuals that caused 
by both the statistical error of the test data and the systematic error of the background model.}
\label{Fig:analysis_err_sys_40th}
\end{figure*}

\begin{figure*}
\center{
\includegraphics[angle=0,scale=1.00]{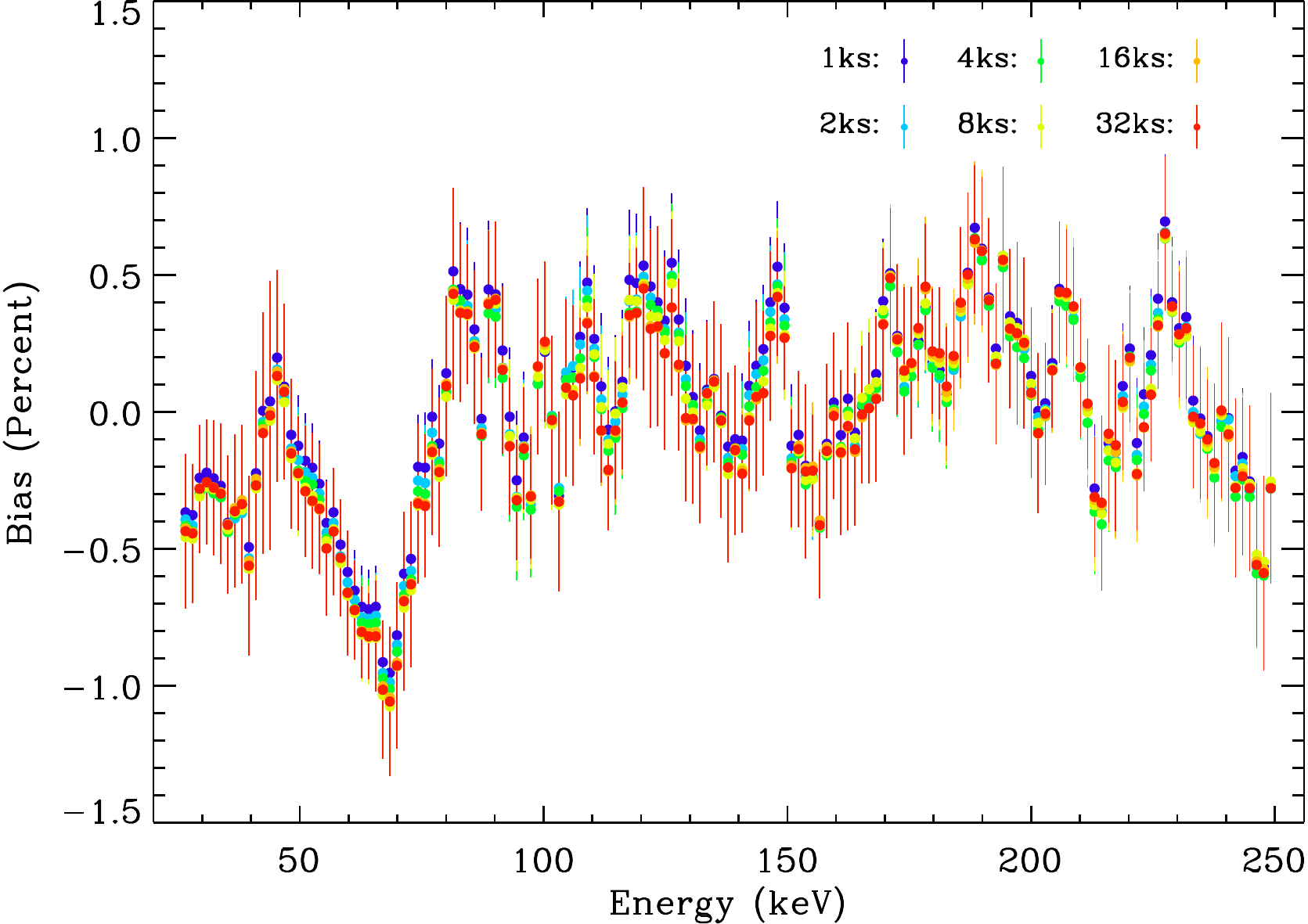}}
\caption{Systematic biases of the background spectra estimation in $26-250$~keV with the exposures 1~ks, 2~ks 4~ks, 8~ks, 16~ks and 32~ks.}
\label{Fig:bias_sys_spec_channel}
\end{figure*}

\begin{figure*}
\center{
\includegraphics[angle=0,scale=1.00]{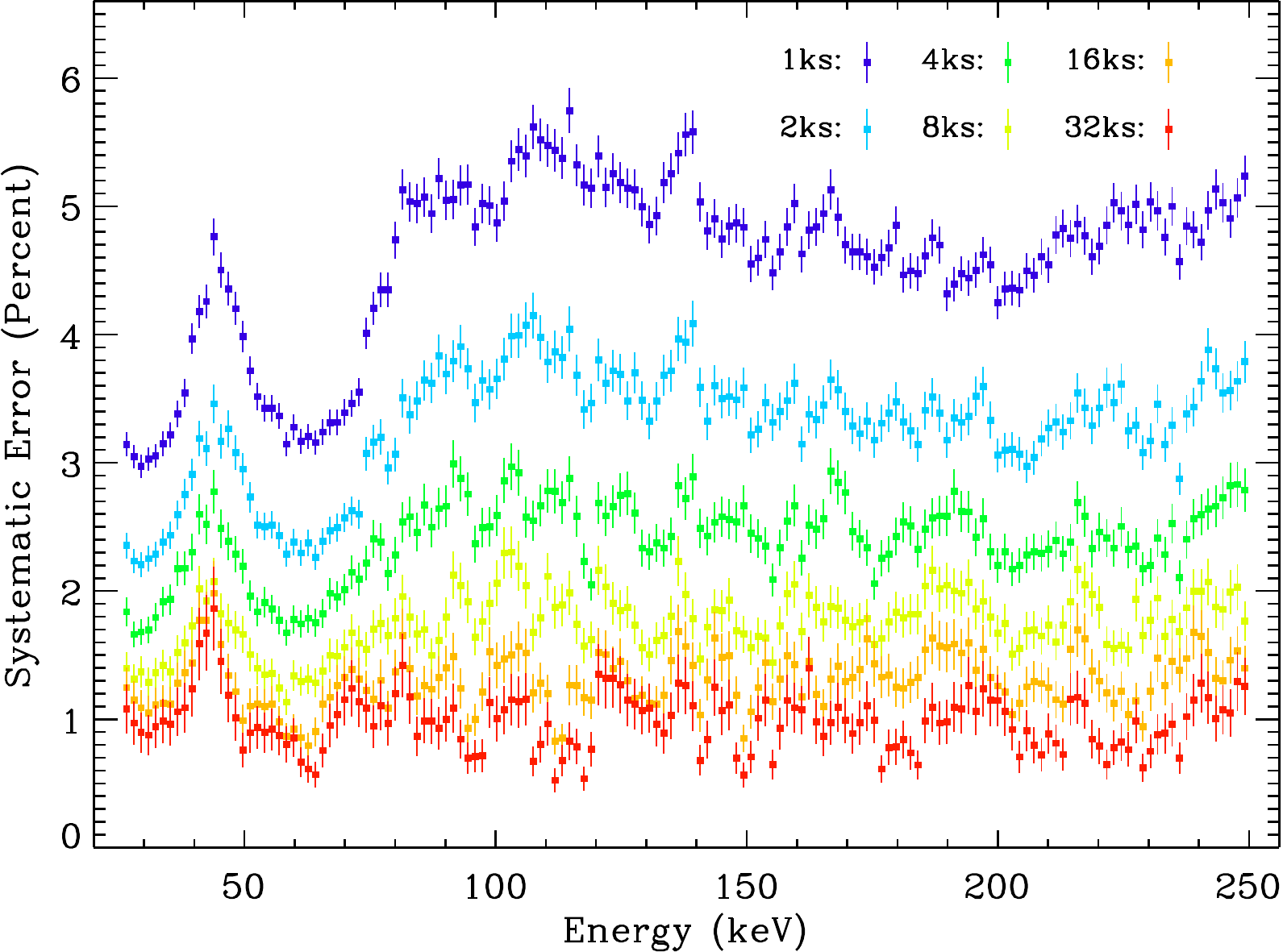}}
\caption{Systematic errors of the background spectra estimation in $26-250$~keV with the exposures 1~ks, 2~ks 4~ks, 8~ks, 16~ks and 32~ks.}
\label{Fig:err_sys_spec_channel}
\end{figure*}

Beside the analysis for $T_{\rm exp}=1~{\rm ks}$, we also perform the same analysis for $T_{\rm exp}=2~{\rm ks}$, 4~ks, 8~ks, 16~ks, 
and 32~ks with the above process to obtain $B$ and $\sigma_{\rm sys}$ of every channel.
As shown in Figure~\ref{Fig:analysis_err_sys_40th}, more than 500 sub-observations with $T_{\rm exp}=1~{\rm ks}$ are obtained, and the residuals $R$ of 
the 24th channel ($E=50$~keV) of each sub-observation fluctuate around zero.
With the method described above, we obtain the bias $B\sim0.1\pm0.2\%$ and the systematic error $\sigma_{\rm sys}\sim4.0\pm0.1\%$.
For the 59th channel ($E=100$~keV) with $T_{\rm exp}=32~{\rm ks}$, 
the bias and systematic error are $B\sim-0.3\pm0.3\%$ and $\sigma_{\rm sys}\sim1.0\pm0.2\%$, respectively. 
Figure~\ref{Fig:bias_sys_spec_channel} and \ref{Fig:err_sys_spec_channel} shows the biases and systematic errors of 
all the channels with $T_{\rm exp}=1~{\rm ks},2~{\rm ks},4~{\rm ks},8~{\rm ks},16~{\rm ks}$, and 32~ks.
The biases are stable with different exposures, whereas the systematic errors decrease with the increasing exposures.
It can be seen that the systematic errors become large relatively in the channel range where the background emission lines are significant.
The average systematic errors in $26-100$~keV for the different exposures are shown in Table~\ref{tab:systematic_error_spectra}.
In further application of the background model, the systematic biases will be corrected.

\subsection{Test for Background Light Curve}
\begin{figure*}
\center{
\includegraphics[angle=0,scale=1.00]{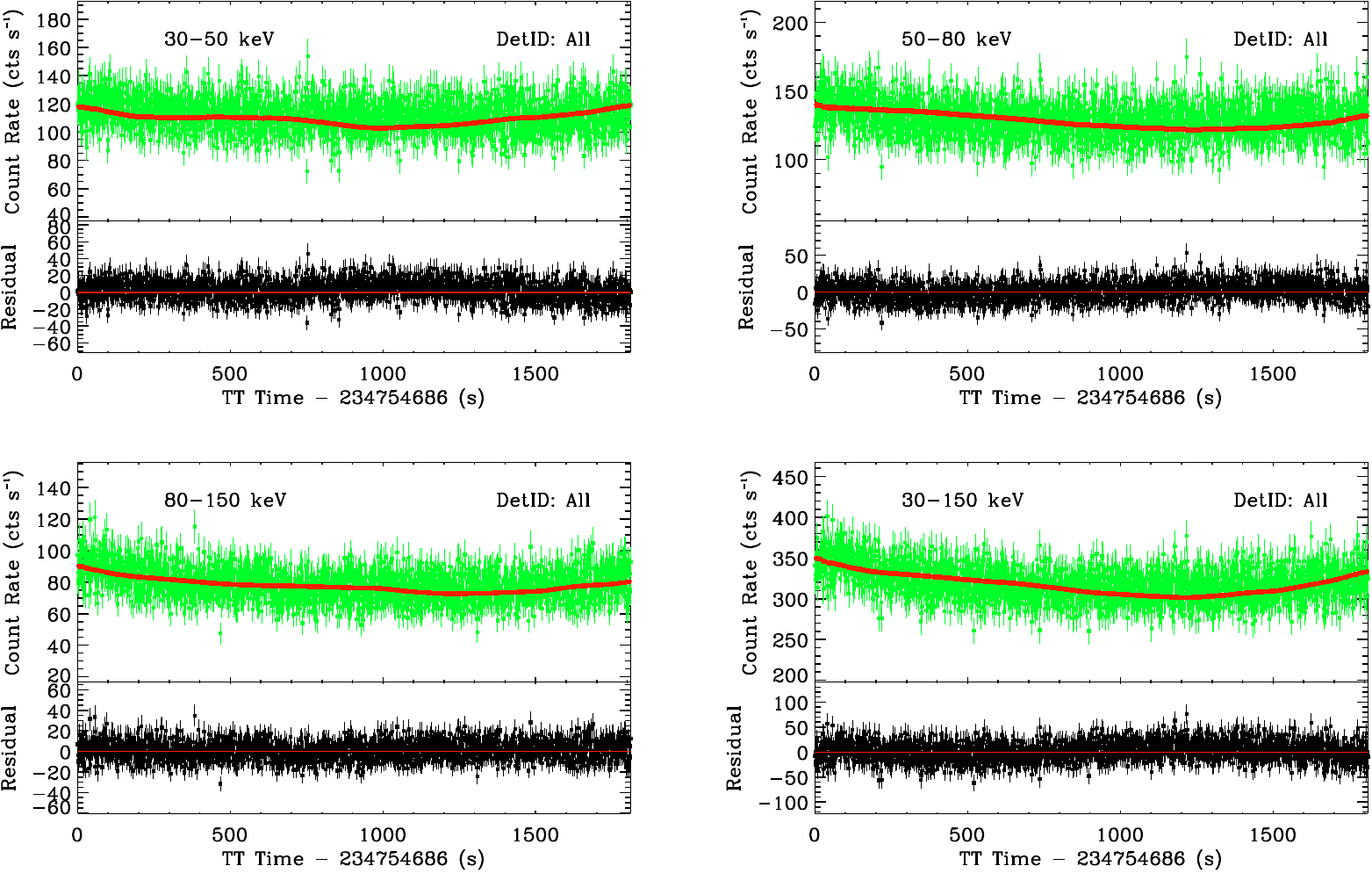}}
\caption{Test of the background light curve estimation in four energy bands with $\rm T_{bin}=1~{\rm s}$.
For each panel, the observed (green) and estimated (red) background light curves are shown in the top, 
and the residual is shown in the bottom.}
\label{Fig:bkg_lc_test}
\end{figure*}

As shown in Figure~\ref{Fig:bkg_lc_test}, the background light curve can be obtained with the process in Section 3.2.
We also do the model test of the background light-curve with different energy bands and different time bins. 
As an example, the model test of the background light curve with energy band $30-150$~keV and $\rm T_{bin}=1~{\rm s}$ is performed by the following steps:

(1) Re-group all the blank sky observations for model test into one observation with the energy band $30-150$~keV and $\rm T_{bin}=1~{\rm s}$;

(2) Obtain the observational light curve ($L_{\rm obs}$) and the background light curve $L_{\rm bkg}$ calculated by the background model;

(3) Make the comparison between $L_{\rm obs}$ and $L_{\rm bkg}$ to obtain the residuals $R_{\rm L} = L_{\rm obs} - L_{\rm bkg}$;

(4) Calculate the mean value and intrinsic dispersion of the residuals with the method in Section 4.1 to obtain bias $B_{\rm L}$ and systematic error $\sigma_{\rm sys, L}$.

\begin{figure*}
\center{
\includegraphics[angle=0,scale=1.00]{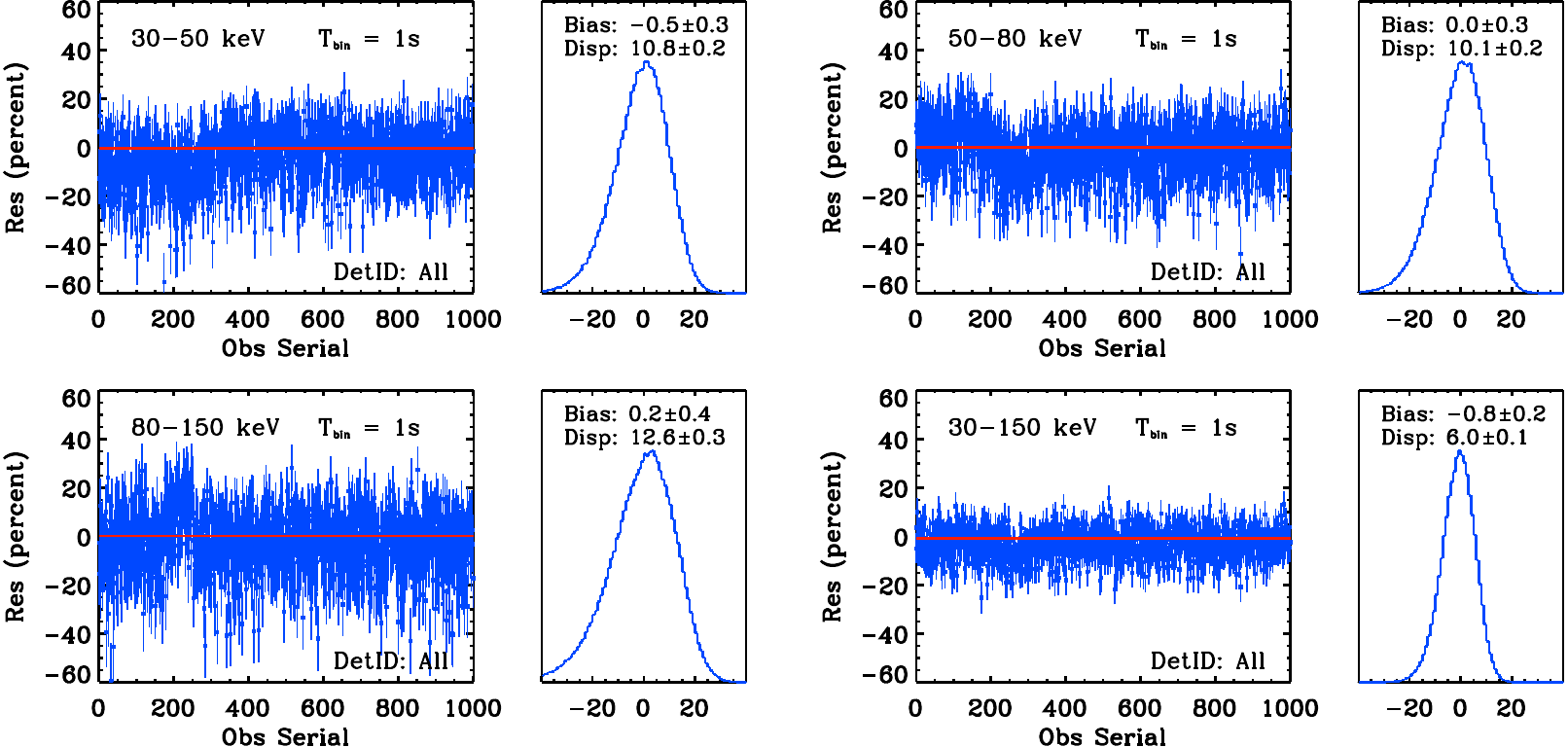}}
\caption{Distributions of the residuals of the background estimations of the light curves with $\rm T_{bin}=1$~s in four energy bands.
In each panel, the broadening of the histogram is the dispersion of the residuals that caused 
by both the statistical error of the test data and the systematic error of the background model.}
\label{Fig:err_sys_lc_1s}
\end{figure*}

\begin{figure*}
\center{
\includegraphics[angle=0,scale=1.00]{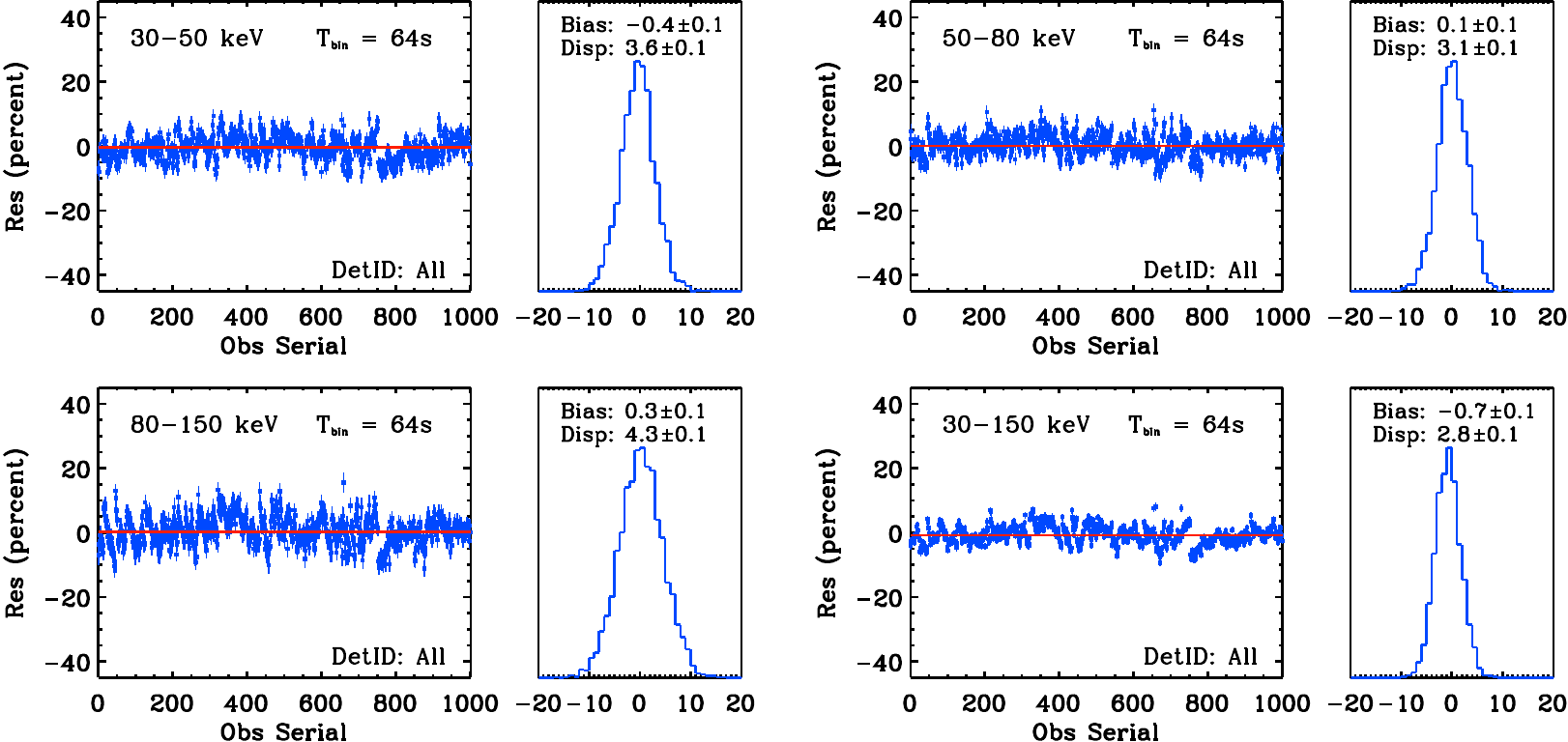}}
\caption{The same as Figure~\ref{Fig:err_sys_lc_1s}, but the time bin of the background light curve is $64$~s.}
\label{Fig:err_sys_lc_64s}
\end{figure*}

We repeat the above steps to test the background light curve with four energy bands and eight time bins.
As shown in Table~\ref{tab:systematic_error_lc}, nearly all the systematic errors are similar for different time bins at each energy band.
For the background light curves with $\rm T_{bin}<16~s$ that are obtained by the interpolation from that with $\rm T_{bin}=16~s$, 
a small time bin can only increase the statistical error of the testing data but can not increase the systematic error of the background estimation.
For the background light curve with $\rm T_{bin}>16~s$, the systematic error is not dominated by the that caused by the propagation
of the statistical error of the blind detector. Thus a large time bin can take a small statistic error of the blind detector 
but can not affect the systematic error significantly.
Figure~\ref{Fig:err_sys_lc_1s}--\ref{Fig:err_sys_lc_64s} shows the residual distribution for four energy bands at two time bins ($\rm T_{bin}=1~{\rm s}$ and 64 s).
It can be seen that all the residuals fluctuate around zero that means the bias of the background estimation is very small 
(e.g., $B_{\rm L}=0.5\%$ for $\rm T_{bin}=1~s$ in $30-50$~keV). 
The detailed result obtained with the above processes are shown in Table~\ref{tab:systematic_error_lc}.
We can see that the systematic errors of the background light curve with different 
energy band and different time bins are all around 3$\%$.

\section{DISCUSSION}
In Section 4, we obtained the systematic errors of the background spectra estimation of \emph{Insight-HXMT}/HE with different exposures. 
Due to various unknown physical processes and inaccuracy of the model assumptions, there exist systematic errors in addition to statistical 
errors. As described in Section 3, the estimation of the background of \emph{Insight-HXMT}/HE has a series of processes. The uncertainties of 
each process can be propagated to the final estimated background, e.g., the selection of empirical functions when fitting the long-term 
evolution of the background, the uncertainty of the assumption the space environment is stable, as well as the effect of satellite attitude on 
the background. In addition, the statistical error of the blind detector with short exposure can be translated into the systematic error of 
the estimated background. Consequently, the systematic error decreases as the exposure increases (Figure~\ref{Fig:err_sys_spec_channel}). At 
present, we use the quadratic polynomial to fit the long-term evolution of the background. As the data becomes more abundant in the future, we 
will try other empirical functions. As the result of the \emph{RXTE}/PCA scanning and slewing observations 
(\citealt{2003A&A...411..329R}), CXB has a fluctuation of 7$\%$ per square deg. According to the FOV of \emph{Insight-HXMT}/HE, it can be 
calculated that CXB has a 2$\%$ fluctuation in \emph{Insight-HXMT}/HE observation. As the proportion of CXB in background is less than 10$\%$, 
therefore the systematic error contributed by the CXB fluctuation is less than 0.2\% and thus can be ignored.

We obtained the systematic errors of the background spectra with different exposures and the background light curves with different time bins. 
For temporal analysis, since the background signal has no special periodicity, the background does not affect the power spectrum in usual 
frequency range, i.e., the influence of background on power spectrum can be ignored (e.g., \citealt{2018ApJ...866..122H}; Xiao et al. 2019a). 
However, the background estimation can affect the fractional rms as the calculation requires the flux of the target. 
\citet{2018ApJ...866..122H} studied the quasi-periodic oscillation of MAXI\,J1535--571, and the fractional rms is calculated by
${\rm rms}=\sqrt{P}(S+B)/S$ (\citealt{2015ApJ...799....2B}). Thus we can find that the systematic error of the background light curve 
can be ignored for the calculation of the fractional rms. For spectral analysis, systematic errors will cause the uncertainty of the spectral 
parameters. Xiao et al. (2019b) studied the cyclotron absorption line of Hercules\,X-1 and obtained accurate parameters of the continuum and 
the cyclotron absorption line. The results are also consistent with the {\sl NuSTAR} result, meaning that means the systematic error has no 
significant influence on the \emph{Insight-HXMT}/HE spectra for bright sources.

The satellite attitude is not considered in the background model described in this paper. The Geant4 simulation on ground shows that 
there exists a weak correlation between the background and satellite altitude. Nevertheless, the influence of satellite attitude on 
the background estimation is much smaller than the current background estimation accuracy. Therefore, even if the satellite attitude is 
considered, it will not significantly improve the accuracy of the background estimation. In the future, as the background estimation is 
more accurate, such that the impact of the satellite attitude on the background estimation cannot be ignored, the relationship between 
satellite attitude and background will be added to the background model to improve the accuracy of the background estimation.

At this stage, the background estimation method is based on the blind detector, which is the unique design of \emph{Insight-HXMT}/HE. 
In addition, the background model independent with the blind detector also will be developed, using the house-keeping parameters, the 
physical principles of the \emph{Insight-HXMT}/HE background, and the experience of the background to obtain a parametric background model, 
which is similar to the {\sl Suzaku}/HXD (\citealt{2009PASJ...61S..17F}).

\section{SUMMARY}
With the large number of blank sky observations of \emph{Insight-HXMT}, we conducted a detailed study of the background of 
\emph{Insight-HXMT}/HE. Both the spectra and the light curves are consistent with the Geant4 simulation on ground. According to the 
characteristics of the background, we have developed a GTI selection criteria recommended for users. The background of \emph{Insight-HXMT}/HE 
is contributed by the particle in orbit and dominated by the long time scale activation background. We first propose and construct the 
background model that combine the background map with long-term evolution and the current observation of blind detectors that is the unique 
design of \emph{Insight-HXMT}. The systematic error analyses of the spectra with different exposures are performed. Currently, the precision 
of the background spectra estimation is better than $2\%$ with the exposure $T_{\rm exp} > 8~{\rm ks}$. The systematic errors of the 
background light curve with different time bins are also analyzed systematically. For the estimated background light curves in $30-150$~keV 
with $\rm T_{bin} = 1$ to $128~{\rm s}$, all the systematic errors are $<3\%$. In the future, the attitude of the satellite will be considered 
and the estimation of \emph{Insight-HXMT}/HE will be more accurate.

\acknowledgments
This work made use of the data from the \emph{Insight-HXMT} mission, a project funded by China National Space Administration (CNSA) 
and the Chinese Academy of Sciences (CAS). The authors thank supports from the National Program on Key Research and Development Project 
(Grant No. 2016YFA0400800) and the National Natural Science Foundation of China under Grants No. U1838202 and U1838201.

\renewcommand{\arraystretch}{1.2}
\begin{deluxetable}{ccccccc}
  \tablewidth{0pt}
  \tablecaption{Average systematic errors of the background spectra ($26-100$~keV) with different exposures}
  \tablehead{  $T_{\rm exp}$  &  1~ks  &  2~ks  &  4~ks  &  8~ks  &  16~ks  &  32~ks  }
  \startdata
      $\sigma_{\rm sys}$  &  3.6\%  &  2.6\%  &  2.0\%  &  1.5\%  &  1.2\%  &  0.9\%
	  \enddata
	  \label{tab:systematic_error_spectra}
\end{deluxetable}

\renewcommand{\arraystretch}{1.2}
\begin{deluxetable}{ccccccccc}
  \tablewidth{0pt}
  \tablecaption{Systematic errors of the background light curves in four energy bands with different time bins}
  \tablehead{  Energy Band  &  1~s  &  2~s  &  4~s  &  8~s  &  16~s  &  32~s  &  64~s  &  128~s  }
  \startdata
	  $30-50$ keV   &  3.1\%  &  2.9\%  &  2.9\%  &  2.8\%  &  2.8\%  &  2.7\%  &  2.6\%  &  2.5\% \\ 
	  $50-80$ keV   &  2.9\%  &  2.6\%  &  2.5\%  &  2.5\%  &  2.4\%  &  2.4\%  &  2.3\%  &  2.2\% \\ 
	  $80-150$ keV  &  4.2\%  &  3.8\%  &  3.7\%  &  3.6\%  &  3.6\%  &  3.5\%  &  3.4\%  &  3.3\% \\ 
	  $30-150$ keV  &  2.6\%  &  2.4\%  &  2.4\%  &  2.4\%  &  2.3\%  &  2.3\%  &  2.2\%  &  2.2\% \\ 
	  \enddata
	  \label{tab:systematic_error_lc}
\end{deluxetable}

\end{document}